\documentclass[fleqn,usenatbib]{mnras} 

\usepackage{newtxtext,newtxmath}

\usepackage[T1]{fontenc}

\DeclareRobustCommand{\VAN}[3]{#2}
\let\VANthebibliography\thebibliography
\def\thebibliography{\DeclareRobustCommand{\VAN}[3]{##3}\VANthebibliography}

\usepackage{graphicx}	
\usepackage{amsmath}	
\usepackage{amssymb}	
\usepackage{xcolor,ulem}

\title[Ejecta Radio Flares Associated with Jets]{Shock within a shock: revisiting the radio flares of NS merger ejecta and GRB-supernovae}

\author[B. Margalit \& T. Piran]{
Ben Margalit$^{1}$\thanks{NASA Einstein Fellow}\thanks{E-mail: benmargalit@berkeley.edu}
and
Tsvi Piran$^{2}$
\\
$^{1}$Astronomy Department and Theoretical Astrophysics Center, University of California, Berkeley, Berkeley, CA 94720, USA\\
$^{2}$Racah Institute of Physics, Edmund J. Safra Campus, Hebrew University of Jerusalem, Jerusalem 91904, Israel\\
}

\date{Accepted XXX. Received YYY; in original form ZZZ}

\pubyear{2015}

\begin{document}
\label{firstpage}
\pagerange{\pageref{firstpage}--\pageref{lastpage}}
\maketitle

\begin{abstract}
Fast ejecta expelled in binary neutron star (NS) mergers or energetic supernovae (SNe) should produce late-time synchrotron radio emission as the ejecta shocks into the surrounding ambient medium. Models for such radio flares typically assume the ejecta expands into an unperturbed interstellar medium (ISM). However, it is also well-known that binary NS mergers and broad-lined Ic SNe can harbor relativistic jetted outflows. In this work, we show that such jets shock the ambient ISM ahead of the ejecta, thus evacuating the medium into which the ejecta subsequently collides. 
Using an idealized spherically-symmetric model,
we illustrate that this inhibits the ejecta radio flare at early times $t < t_{\rm col} \approx 12 \, {\rm yr} \, (E_{\rm j}/10^{49} \, {\rm erg})^{1/3} (n/1 \, {\rm cm}^{-3})^{-1/3} (v_{\rm ej}/0.1c)^{-5/3}$ where $E_{\rm j}$ is the jet energy, $n$ the ISM density, and $v_{\rm ej}$ the ejecta velocity.
We also show that this
can produce a sharply peaked enhancement in the light-curve at $t = t_{\rm col}$. 
This has implications for radio observations of GW170817 and future binary NS mergers, gamma-ray burst (GRB) SNe, decade-long radio transients such as FIRST J1419, and possibly other events where a relativistic outflow precedes a slower-moving ejecta. Future numerical work will extend these analytic estimates and treat the multi-dimensional nature of the problem.
\end{abstract}

\begin{keywords}
gamma-ray bursts --
neutron star mergers --
transients: supernovae --
shock waves --
radiation mechanisms: non-thermal --
radio continuum: transients
\end{keywords}



\section{Introduction}

Shocks are ubiquitous phenomena in astrophysical settings and are well-studied sites of magnetic field amplification and non-thermal particle acceleration \citep{Bell78,Blandford&Eichler87}. Consequently, shocks are capable of producing bright synchrotron emission detectable across the electromagnetic spectrum. Models of synchrotron shock emission have been extremely successful in explaining observations of radio supernovae (SNe; \citealt{Chevalier82,Chevalier1998}), gamma-ray burst (GRB) afterglows \citep{Paczynski&Rhoads93,Meszaros&Rees97,Rhoads97,Sari+98,Rhoads99} and other events associated with fast outflows and/or occurring in dense environments (see e.g. Fig.~5 of \citealt{Margutti+18}).

Binary neutron star (BNS) mergers can expel $\sim 10^{-3}-10^{-1} M_\odot$ of material at a fraction of the speed-of-light \citep{Lattimer&Schramm74,Lattimer&Schramm76,Rosswog+99,Bauswein+13,Radice+16,Siegel+17}. Radioactive decay of freshly-synthesized $r$-process elements in this ejected material can subsequently power an optical/near-infrared kilonova \citep[also known as macronova;][]{Li&Paczynski98,Kulkarni05,Metzger+10,Barnes&Kasen13,Tanaka&Hotokezaka13}, such as has been recently observed in conjunction with the first gravitational-wave (GW) detected BNS merger, GW170817 \citep{Coulter+17,Cowperthwaite+17,Kasliwal+17,Tanvir+17,LVC_multimessenger}.

Similarly to shocks in radio SNe,
the expanding kilonova ejecta has been predicted to produce a $\sim$decade-long synchrotron radio flare as the ejecta shocks the ambient interstellar medium (ISM; \citealt{Nakar&Piran11}).
The work of \cite{Nakar&Piran11} has been extended by various authors to account for more realistic kilonova-ejecta distributions \citep[e.g.][]{Piran+13,Margalit&Piran15,Hotokezaka&Piran15} with qualitatively similar results, and has been applied to GW170817 in attempt to predict it's radio-flare signature and observationally constrain the ejecta parameters \citep[e.g.][]{Alexander+17,Kathirgamaraju+19,Hajela+19}. It has also been used to place limits on ejecta radio flares that would be expected to accompany short GRBs \citep{Metzger&Bower14,Horesh+16,Fong+16,Klose+19}.

\begin{figure*}
    \centering
    \includegraphics[width=0.95\textwidth]{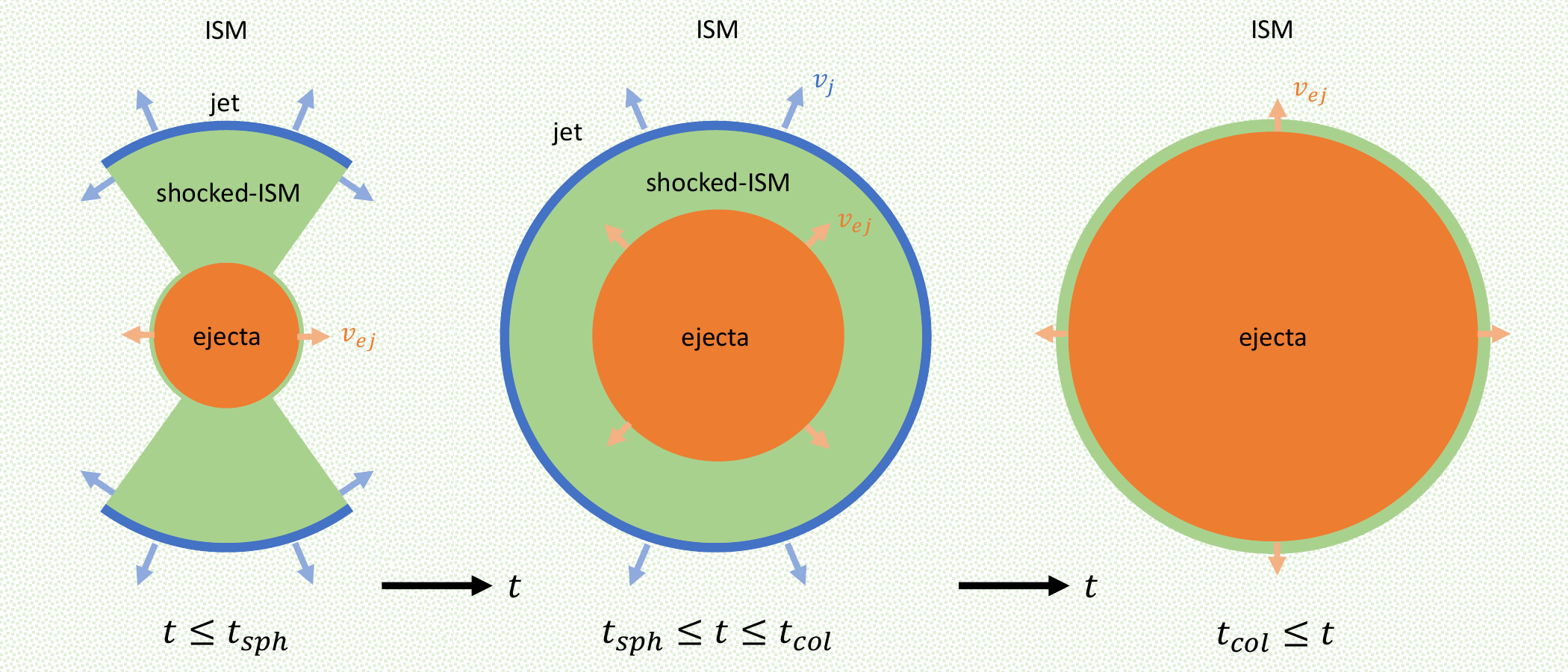}
    \caption{Cartoon illustration of the two components considered in this work --- an initially ($t < t_{\rm sph}$; eq.~\ref{eq:tsph}) collimated ultra-relativistic jet (blue) and sub-relativistic spherical ejecta (orange) that expand simultaneously into an ambient ISM (green). The jet decelerates and spreads azimuthally as it sweeps-up an increasing ISM mass, until, at $t \gtrsim t_{\rm sph},t_{\rm NR}$, the jet has become sub-relativistic and quasi-spherical. At this point, the jet forward-shock expanding at $v=v_{\rm j}$ is well described by the Sedov-Taylor solution (eq.~\ref{eq:Rjvj}), and the ejecta which begins catching up with the jet forward shock propagates within the cavity of jet-shocked ISM. At time $t=t_{\rm col}$ the ejecta overtakes the jet-ISM forward shock (eq.~\ref{eq:tcol}) and subsequently expands into an unperturbed ISM.}
    \label{fig:cartoon}
\end{figure*}

Here we point out a shortcoming in these models that has been previously overlooked. Ejecta radio-flare models in the literature typically assume that the kilonova ejecta is expanding into a cold unperturbed (constant-density, static) ISM, however --- BNS merger events like GW170817 produce an ultra-relativistic collimated jet \citep{Mooley+18b} that precedes the kilonova ejecta. This jet runs into and shocks the surrounding ISM before the ejecta, therefore changing the medium with which the ejecta subsequently collides (Fig.~\ref{fig:cartoon}). 
In the following we explore the effect that this ``preshaping'' of the ambient medium ahead of the ejecta has on the predicted ejecta radio-flare.

The situation described above for BNS mergers is also relevant in other astrophysical settings as well. A completely analogous situation occurs for long GRBs, which are known to be accompanied by energetic broad-lined Ic SNe \citep{Galama+98,Bloom+99,Hjorth+03,Woosley&Bloom06}.
The late-time radio emission from interaction of these energetic SNe ejecta with the surrounding ISM has been studied by \cite{BarniolDuran&Giannios15} and subsequently extended and applied to observations \citep{Kathirgamaraju+16,Peters+19}. Such studies however similarly have yet to account for the interaction between the GRB-jet and SN-ejecta, and have instead treated these as independent components propagating into an unperturbed ISM. In \S\ref{subsec:LGRB_IcSNe} we apply our results to this problem and show that this interaction can affect the predicted light-curves and implied source-parameter constraints.

This paper is structured as follows: we begin by reviewing the standard results for jet afterglows and ejecta radio-flares produced when these run independent of one another into an unperturbed constant-density ISM (\S\ref{sec:non-interacting}). In \S\ref{sec:interaction} we extend these results and calculate the affect that interaction between these two components has on the resulting light-curve. We then discuss implications of our results in application to various astrophysical settings (\S\ref{sec:Implications}), and conclude with a discussion of caveats, observational implications, and directions for future work (\S\ref{sec:Discussion}).
We focus on an ISM cirum-stellar medium in the main text, but generalize our results to a wind environment in Appendix~\ref{sec:Appendix_winddensity}.

\section{Non-Interacting Jet, Ejecta}
\label{sec:non-interacting}

We begin by discussing the relevant timescales and standard results for jet afterglow and ejecta radio flares expanding into a cold, constant-density ISM.
Much of this has been previously discussed by \cite{BarniolDuran&Giannios15} and references therein. In the following, we repeat and summarize the main results for completeness.

\subsection{Dynamics and timescales}

From dimensional arguments alone, a characteristic timescale for a relativistic jet of total energy\footnote{Note that we consider here the total energy of the jet and not the isotropic equivalent energy as commonly used in the afterglow literature.} $E_{\rm j}$ propagating into an ISM of number density $n$ is
\begin{equation}
\label{eq:tauj}
    \tau_{\rm j} = \left(\frac{E_{\rm j}}{n m_p c^5}\right)^{1/3}
    \approx 0.20 \, {\rm yr} \, E_{{\rm j},49}^{1/3} n_0^{-1/3} .
\end{equation}
Above and in the following we adopt the notation $Q_x \equiv Q / 10^x$ for any quantity $Q$ in appropriate cgs units.

A collimated relativistic jet of half-opening angle $\theta_{\rm j}$ and Lorentz factor $\Gamma \gg \theta_{\rm j}^{-1}$ will initially propagate radially without appreciable lateral expansion since there is no causal contact between different regions of the jet. In typical astrophysical scenarios the jet is double-sided, i.e. there will be two antipodally-symmetric jets (see Fig.~\ref{fig:cartoon}), and we assume here that both have identical half-opening angles $\theta_{\rm j}$ and energy $E_{\rm j}/2$ (defined this way, $E_{\rm j}$ is the total energy of the bipolar outflow).
While the jet is ultra-relativistic and collimated it's evolution is well described by the Blandford-McKee solution with the isotropic-equivalent energy $E_{\rm j,iso} = E_{\rm j} / (1 - \cos{\theta_{\rm j}}) \approx 2 E_{\rm j} / \theta_{\rm j}^2$ \citep{Blandford&McKee76}.

This breaks down once the jet Lorentz factor drops to $\Gamma \sim \theta_{\rm j}^{-1}$, at which point different regions in the jet come into causal contact and azimuthal spreading can commence. The time at which this occurs is known as the jet-break time, which, measured in the lab frame is \citep{Sari+99}
\begin{equation}
\label{eq:tj}
    t_{\rm jb} = \left( \frac{17 E_{\rm j,iso}}{16\pi n m_p c^5 \theta_{\rm j}^{-2}} \right)^{1/3} 
    = \left(\frac{17}{8\pi}\right)^{1/3} \tau_{\rm j} .
\end{equation}
In the observer frame this is measured to occur earlier by a factor $1/\Gamma(t_{\rm jb})^2 \sim \theta_{\rm j}^2 \ll 1$.\footnote{Here and in the following derived expressions we do not include explicit dependence on cosmological redshift $\propto (1+z)^{-1}$.}

Following the jet-break time (eq.~\ref{eq:tj}) the jet can start spreading azimuthally.
The detailed dynamics of this jet-spreading phase have been investigated by many works \citep{Rhoads99,Granot+01,Ayal2001,Zhang&MacFadyen09,Wygoda+11,DeColle+12,Granot&Piran12,Duffell&Laskar18}.
Initial analytic models suggested a fast exponential phase where the jet sphericizes at $t \sim t_{\rm jb}$ \citep{Rhoads99}, however later numerical simulations and semi-analytic models indicate that jet spreading
is delayed and only begins in earnest once the (still-collimated) jet decelerates to non-relativistic velocities at $t=t_{\rm NR}$ \citep{DeColle+12,Granot&Piran12}. This time is defined by the Sedov-Taylor timescale of the isotropic-equivalent one-sided jet energy $E_{\rm j,iso}$. Thus, the initially-collimated jet becomes quasi-spherical\footnote{Early work by \cite{Ayal2001} showed that the outflow will become fully spherical on a time scale of few thousand years (see also \citealt{Ramirez-Ruiz&MacFadyen10}). However, quasi-sphericity is in fact reached much earlier, on the  timescale that we estimate here \citep{Zhang&MacFadyen09}.
}
and approaches the spherical Sedov-Taylor solution for an explosion of energy $E_{\rm j}$ after time $t=t_{\rm sph}$, where
\begin{align}
\label{eq:tsph}
    t_{\rm sph} \gtrsim t_{\rm NR} 
    &\sim \left( \frac{3 E_{\rm j,iso}}{4\pi n m_p c^5} \right)^{1/3} 
    = \left(\frac{3}{2\pi}\right)^{1/3} \theta_{\rm j}^{-2/3} \tau_{\rm j}
    \\ \nonumber
    &\approx 
    0.73 \, {\rm yr} \, E_{{\rm j},49}^{1/3} n_0^{-1/3} \theta_{{\rm j},-1}^{-2/3} \    .
\end{align}
Note that $t_{\rm sph}$ can exceed $t_{\rm NR}$ by a non-negligible factor $\sim 3-10$ and that the exact evolution can only be fully explored numerically.

At times $t \gg t_{\rm sph},t_{\rm NR}$ the ``jet'' is non-relativistic and spherical and is therefore well-described by the Sedov-Taylor solution. We continue to term this component the `jet' even though at such times the outflow has lost it's initially jet-like features (strong collimation and ultra-relativistic velocities). This is to distinguish it from the matter-dominated ejecta, which is also sub-relativistic and quasi-spherical.
The jet-ISM forward shock position at times $t \gg t_{\rm sph}, t_{\rm NR}$ are thus
\begin{equation}
\label{eq:Rjvj}
    R_{\rm j} = 
    \xi
    \left(\frac{E_{\rm j}}{n m_p}\right)^{1/5} t^{2/5}; ~~~~~~~~~~
    v_{\rm j} = \frac{2}{5} \frac{R_{\rm j}(t)}{t} ,
\end{equation}
where the numerical constant 
$\xi \simeq 1.17$
is determined by the Sedov-Taylor solution for an adiabatic index of $g=5/3$ \citep{Sedov59}. Barring any other interaction, the jet remains in the Sedov-Taylor phase till radiative losses at the shock front become energetically important. This does not occur on timescales of interest for this problem, and therefore we do not discuss subsequent evolution phases of the jet-ISM shock.

The bulk ejecta of SNe or BNS mergers are launched quasi-spherically at sub-relativistic velocities $\beta_{\rm ej} = v_{\rm ej}/c < 1$.
Although different launching mechanisms likely contribute to different angular regions of BNS merger kilonova ejecta (e.g. equatorial tidal ejecta vs. polar disk winds and shock outflows; see \citealt{Shibata&Hotokezaka19} for a recent review), the global structure of the kilonova ejecta remains quasi-spherical, especially in cases where both polar and equatorial outflows are present (if the remnant does not promptly collapse to a black hole; see e.g. \citealt{Margalit&Metzger19}). Deposition of thermal energy through radioactive decay of $r$-process material in the ejecta acts to further drive the ejecta towards spherical-symmetry \citep{Grossman+14}. Although the oblate nature of this ejecta \citep[e.g.][]{Radice+18} can impact some details of the afterglow emission \citep{Margalit&Piran15}, the overall light-curve is reasonably approximated by spherically-symmetric models. In the following we therefore assume a spherically-symmetric ejecta.

A further complication arises from the velocity-structure of SNe or kilonova ejecta. Even for spherical ejecta, the outflow is expected to have a distribution of velocities with a small, but potentially important, amount of material moving at velocities much greater than the ``bulk'' velocity of the ejecta \citep[e.g.][]{Chevalier&Soker89,Matzner&McKee99,Nakar&Piran11}.
In the following we neglect these details and treat the simplified case of a so-called `single-velocity shell' \citep[see][]{Margalit&Piran15} --- approximating the ejecta as an outflow with kinetic energy $E_{\rm ej}$ expanding with a single bulk velocity $v_{\rm ej}$. This captures the qualitative features of the problem and is also accurate at late times ($\gg t_{\rm dec}$; eq.~\ref{eq:tdec}), however underestimates the ejecta radio-flare luminosity at early times.
In Appendix~\ref{sec:Appendix_velocityprofile} we extend some of the results of the following sections to an ejecta with a velocity structure, still within the framework of a non-relativistic spherical problem. For BNS mergers, the outflow velocity profile is intricately related to its non-spherical structure, a problem we leave for investigation in future numerical work.

A massive ejecta expanding at velocity $v_{\rm ej}$ into a cold constant-density ISM will initially coast at constant velocity $v_{\rm ej}$ ($R_{\rm ej}=v_{\rm ej}t$). This will change and the ejecta become affected by the ambient medium once it has swept up an ISM mass comparable to its own ($\sim E_{\rm ej}/v_{\rm ej}^2$). The characteristic time for this to occur is 
\begin{equation}
\label{eq:tdec}
    t_{\rm dec} = \left( \frac{3 E_{\rm ej}}{4\pi n m_p v_{\rm ej}^5} \right)^{1/3}
    \approx 27 \, {\rm yr} \, E_{{\rm ej},51}^{1/3} n_0^{-1/3} \beta_{{\rm ej},-1}^{-5/3}
\end{equation}{}
known as the deceleration, or Sedov-Taylor, timescale. The latter stems from the fact that, at $t \gg t_{\rm dec}$ the ejecta asymptotes to the Sedov-Taylor solution (i.e. the initial mass of the ejecta is negligible compared to the swept-up ISM mass, and therefore the shock is describable as a point-explosion of energy $E_{\rm ej}$).
\cite{Nakar&Piran11} showed that at typical $\sim$GHz radio frequencies, the radio flare of sub-relativistic ejecta peak at $t=t_{\rm dec}$.

\subsection{Radio light-curves}

\begin{figure*}
    \centering
    \includegraphics[width=0.95\textwidth]{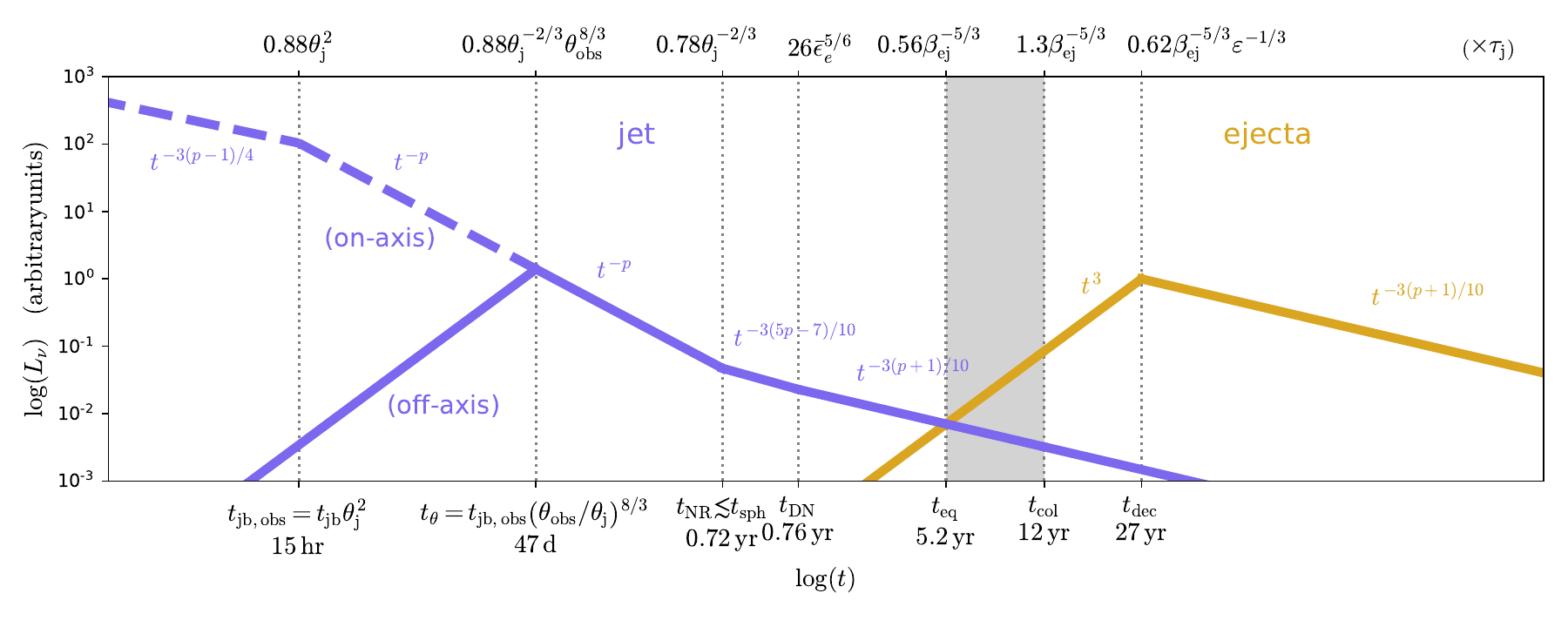}
    \caption{
    Schematic jet-afterglow (blue) and ejecta radio-flare (yellow) light-curves {\it assuming these are independent of one-another} (i.e. neglecting interaction between the two outflow components and instead assuming that both propagate into a cold constant-density ISM).
    The regime of focus in this present work is $t_{\rm eq} < t < t_{\rm col}$ (shaded region), where we predict deviations from the standard light-curve illustrated above; a strong suppression at first and a bright peak towards the end  (see Fig.~\ref{fig:numerical_results} and \S\ref{sec:interaction}). The top horizontal axis presents the hierarchy of timescales measured in the observer frame with respect to $\tau_{\rm j} \approx 0.2 \, {\rm yr} \, E_{{\rm j},49}^{1/3} n_0^{-1/3}$ (eq.~\ref{eq:tauj}), and as a function of the dimensionless parameters of the problem: the ratio of jet to ejecta energy $\varepsilon \equiv E_{\rm j}/E_{\rm ej}$, the ejecta velocity $\beta_{\rm ej} = v_{\rm ej}/c$, shock acceleration microphysical parameters $\bar{\epsilon}_e \equiv 4 \epsilon_e (p-2)/(p-1)$, the initial jet opening angle $\theta_{\rm j}$, and the angle between the observer and jet axis $\theta_{\rm obs}$. These timescales are described in the main text. 
    The time values labeled at the very bottom are calculated for a fiducial set of parameters ($E_{\rm j}=10^{49} \, {\rm erg}, n=1 \, {\rm cm}^{-3}, \theta_{\rm j}=0.1, \theta_{\rm obs}=0.5, \bar{\epsilon}_e=0.1, \beta_{\rm ej}=0.1, E_{\rm ej}=10^{51} \, {\rm erg}$) and can change dramatically for different parameters.
    Temporal scalings for different portions of the light-curve are also listed \citep{Sari+99,Frail+00,Nakar&Piran11,Sironi&Giannios13}. The dashed portion of the blue curve shows the light-curve for an on-axis observer, while the solid is for an off-axis observer. The light-curve in the latter case peaks once the jet Lorentz factor decelerates to $\Gamma \sim \theta_{\rm obs}^{-1}$, which occurs at time $t_\theta = t_{\rm jb,obs} (\theta_{\rm obs}/\theta_{\rm j})^{8/3}$ in the observer frame.}
    \label{fig:schematic}
\end{figure*}{}

Having discussed in the previous section the dynamics of outflows propagating into a cold constant-density ISM, we are now in a position to outline relevant results regarding the radio synchrotron emission produced by such interaction \citep[see][and references therein for further details]{Nakar&Piran11}.

Synchrotron emission from non-relativistic shocks was first discussed in the radio-SN literature \citep[e.g.][]{Chevalier82,Chevalier1998}. Here we follow the formalism of \cite{Sironi&Giannios13} that was developed in the context of GRB afterglows.
One pertinent point is the fact that, at times of interest, both jet and ejecta are marginally within the `deep-Newtonian regime' discussed by \cite{Sironi&Giannios13}. When the velocity of the forward-shock propagating into the ISM is sub-relativistic the bulk of electrons are not accelerated to relativistic velocities and this has to be taken into account in the shock accelerated synchroton emission estimates. 
This regime commences   once the shock velocity drops below
\begin{equation}
    v \lesssim v_{\rm DN} = \left( 8\frac{m_e}{m_p} \right)^{1/2} \bar{\epsilon}_e^{-1/2} c \approx 0.21 c \, \bar{\epsilon}_{e,-1}^{-1/2} ,
\end{equation}
where $\bar{\epsilon}_e \equiv 4 \epsilon_e (p-2)/(p-1)$, $2<p<3$ is the power-law index of non-thermal electrons accelerated at the shock front, and $\epsilon_e$ is the fractional shock power diverted to this non-thermal electron population. 
Synchrotron afterglow emission at times $t<t_{\rm DN}$ (when $v>v_{\rm DN}$) peaks at the frequency of emitting electrons with Lorentz factor $\gamma \sim \gamma_m \gg 1$, while in the deep-Newtonian regime ($t>t_{\rm DN}$) emission is dominated by electrons with $\gamma \sim 2$. This causes the afterglow light-curve to decay less steeply with time following the deep-Newtonian transition \citep{Sironi&Giannios13}.

At frequencies $\nu > \nu_m$ of electrons with Lorentz factors $\gamma_m$, the optically-thin synchrotron luminosity is $L_\nu \propto \gamma \left(dN/d\gamma\right) B \propto \gamma_m^{p-1} B^{(p+1)/2} N$ where $B \propto n^{1/2} v$ is the post-shock amplified magnetic field, and $N \propto n R^3$ the total number of radiating electrons. At early times ($t<t_{\rm DN}$) $\gamma_m \propto v^2$, while in the deep-Newtonian regime $\gamma_{m,{\rm eff}} \sim 2$ and the effective number of relativistic electrons with $\gamma > \gamma_m$ is reduced by a factor $\zeta \propto v^2$, so that $N \to \zeta N$ \citep{Sironi&Giannios13}.
The synchrotron luminosity thus scales as
\begin{align}
\label{eq:Lnu_scaling}
    L_\nu &\propto 
    n^\frac{p+5}{4} R^3
    \begin{cases}{}
    v^\frac{5p-3}{2}; 
    &t<t_{\rm DN}
    \\
    v^\frac{p+5}{2}; &t>t_{\rm DN}
    \end{cases}
    .
\end{align}
At times $t<t_{\rm dec}$ the ejecta coasts at a constant velocity and therefore its luminosity scales as $L_\nu \propto t^3$, whereas in the Sedov-Taylor phase --- that is at $t>t_{\rm sph},t_{\rm NR}$ for the jet and $t>t_{\rm dec}$ for the ejecta --- the radio synchrotron luminosity of these outflows scales as $L_\nu \propto t^{-3(p+1)/10}$ in the deep-Newtonian regime and $L_\nu \propto t^{-3(5p-7)/10}$  otherwise \citep{Frail+00,Nakar&Piran11,Sironi&Giannios13}.

From eq.~(\ref{eq:Lnu_scaling}) we can estimate the time $t_{\rm eq}$ at which, {\it neglecting any interaction between jet and ejecta}, the two outflows' radio emission would equal one another. Since by this assumption, both ejecta and jet propagate into the same cold constant-density ISM, the factor depending on $n$ in eq.~(\ref{eq:Lnu_scaling}) is identical for both. At times $t_{\rm sph} < t < t_{\rm dec}$ however, the dynamics of these components differ. Using eq.~(\ref{eq:Rjvj}) along with $R_{\rm ej}=v_{\rm ej}t$ we find
\begin{align}
\label{eq:teq}
    t_{\rm eq} 
    &= 
    \xi^{5/3} 
    \beta_{\rm ej}^{-5/3} \tau_{\rm j}
    \begin{cases}
    \left(\frac{2}{5}\right)^\frac{5(5p-3)}{3(5p+3)}; &t_{\rm eq}<t_{\rm DN}
    \\
    \left(\frac{2}{5}\right)^\frac{5(p+5)}{3(p+11)}; &t_{\rm eq}>t_{\rm DN}
    \end{cases}
    \\ \nonumber
    &\approx
    5 \, {\rm yr} \, E_{{\rm j},49}^{1/3} n_0^{-1/3} \beta_{{\rm ej},-1}^{-5/3} \     ,
\end{align}
where the fiducial value in the last line is estimated for $2 \leq p \leq 2.7$. This result is consistent within factors of order unity with the previously derived result by \cite{BarniolDuran&Giannios15}.\footnote{Note that the expression for $t_{\rm eq}$ found by these authors does not explicitly depend on $p$. This dependence formally arises due to the differing velocity of jet and ejecta at fixed radius (by a factor of $2/5$; eq.~\ref{eq:Rjvj}), though quantitatively the result depends only weakly on the value of $p$.}

Figure~\ref{fig:schematic} shows the schematic radio synchrotron light-curves both for a jet afterglow (blue) and an ejecta radio flare (yellow), illustrating the hierarchy of timescales and temporal scalings described in this section. Note that in this figure, as in the above discussion, we have assumed that the observing frequency $\nu$ satisfies $\nu_{\rm ssa}, \nu_m < \nu < \nu_{\rm c}$ where $\nu_{\rm ssa}$ is the synchrotron self-absorption frequency and $\nu_{\rm c}$ the cooling frequency. This regime is quite generic in the radio band and at times of relevance \citep[see][for further details]{Nakar&Piran11}, however this may not necessarily apply in the early GRB afterglow phase in which case the light-curve temporal scaling at these times ($\ll t_{\rm NR}$) may differ from the quoted values in this schematic figure \citep[see][for a full discussion of the various scenarios]{Sari+98}.

\section{Jet-Ejecta Interaction}
\label{sec:interaction}

In the previous section we recapitulated results for systems that have either a jet or an ejecta expanding into a cold constant density ISM. We now turn to address systems in which both jet- and ejecta-like outflows are expected. In such systems the ejecta radio-flare will differ from the standard results summarized in the previous section. This is because, at early times, the ejecta does not expand into a cold constant-density ISM. Rather, it propagates into a hot `bubble' of material that has been pre-shocked by the preceding jet-ISM forward shock (Fig.~\ref{fig:cartoon}). This is relevant as long as the jet-ISM forward shock in fact precedes the ejecta (which occurs because the jet initially propagates much faster than the non-relativistic ejecta).
Assuming that the ejecta, expanding within the ``cavity'' created by the jet forward shock, is unaffected by the dilute medium in this cavity, it will coast at constant velocity $v_{\rm ej}$ while $R_{\rm ej} < R_{\rm j}$. This assumption is reasonable so long as $E_{\rm ej} > E_{\rm j}$, which is the case for astrophysical sources of interest.
In this limit, we can easily find the time $t_{\rm col}$ at which the ejecta will overtake the jet forward-shock and collide with the swept-up shell of jet-shocked ISM. From eq.~(\ref{eq:Rjvj}) we find that $v_{\rm ej} t_{\rm col} = R_{\rm j}(t_{\rm col})$ at
\begin{equation}
\label{eq:tcol}
    t_{\rm col} = 
    \xi^{5/3} 
    \beta_{\rm ej}^{-5/3} \tau_{\rm j}
    \approx 12 \, {\rm yr} \, E_{{\rm j},49}^{1/3} n_0^{-1/3} \beta_{{\rm ej},-1}^{-5/3}
    .
\end{equation}
Note that, conveniently, $t_{\rm eq}$ (eq.~\ref{eq:teq}) scales with $t_{\rm col}$, and for any $2 \leq p \leq 3$ we find that $t_{\rm eq}/t_{\rm col} \sim 0.4$. The collision time is also simply related to the decceleration time of the ejecta (eq.~\ref{eq:tdec}) as $t_{\rm col} \approx 2.04 (E_{\rm j} / E_{\rm ej})^{1/3} t_{\rm dec}$.

At times $t>t_{\rm col}$ the ejecta has overtaken the jet forward shock and indeed runs into a cold unperturbed ISM. It is also energetically dominant at these times and so the radio flare signature at $t>t_{\rm col}$ reverts to the standard picture \citep{BarniolDuran&Giannios15}. At times $t<t_{\rm col}$ however, this description is no longer valid since the jet shocks the ISM ahead of the ejecta.


This pre-shocking of the ISM by the jet affects the early-time ejecta radio flare by: (i) changing the density distribution into which the ejecta expands --- sweeping-up the ISM into a $\sim$thin shell at $R_{\rm j}$ and creating a dilute cavity with rising density profile interior to it; 
(ii) giving the shocked-ISM an outwards bulk velocity so that the relative velocity between the ejecta and swept-up material is reduced; 
and (iii) heating the post-shock gas to high temperatures. 
All three can act to dramatically inhibit the early afterglow signal of the ejecta at times $t \ll t_{\rm col}$. However, as we show later in this section --- a sharp peak in the light-curve rising above the naive unperturbed-ISM predictions is expected when the ejecta passes through the thin shell of swept-up ISM at $t=t_{\rm col}$.

In the limit where the jet forward-shock sweeps the surrounding ISM into an infinitesimally thin shell, the ejecta does not produce any snychrotron emission whatsoever before colliding with this swept-up shell. This is qualitatively consistent with what we find in our numerical results, which indicate that the ejecta radio-flare light-curve rises sharply only close to $t_{\rm col}$.
We can estimate this from the Sedov-Taylor solution.

The density profile implied by the Sedov-Taylor solution far interior to the forward shock, $r \ll R_{\rm j}$, is approximately
$n(r \ll R_{\rm j}) \propto r^{3/(g-1)} t^{-6/5(g-1)}$, where $g$ is the gas adiabatic index.
In the same regime, the post-shock gas velocity is simply $v(r \ll R_{\rm j}) \propto r / t$.
Thus, the ejecta synchrotron radio light-curve produced by interaction with this post-shock medium would rise sharply with time,
\begin{align}
\label{eq:Lnu_ej_tscaling}
    L_{\nu,{\rm ej}}\left(t<t_{\rm col}\right) 
    &\propto
    n\left(R_{\rm ej}(t),t\right)^\frac{p+5}{4} R_{\rm ej}(t)^3
    \left[ v_{\rm ej} - v\left(R_{\rm ej}(t),t\right) \right]^a
    \nonumber
    \\
    &\underset{t \ll t_{\rm col}}{\propto} t^{\frac{3(15p+91)}{40}}
\end{align}{}
where the exponent $a$ depends on whether the shock velocity is in the deep-Newtonian regime or not (eq.~\ref{eq:Lnu_scaling}), however the result does not depend on the value of $a$ because the shock velocity (term in square brackets) does not vary with time. In the final line we have taken $g=5/3$ for the adiabatic index.

\begin{figure}
    \centering
    \includegraphics[width=0.45\textwidth]{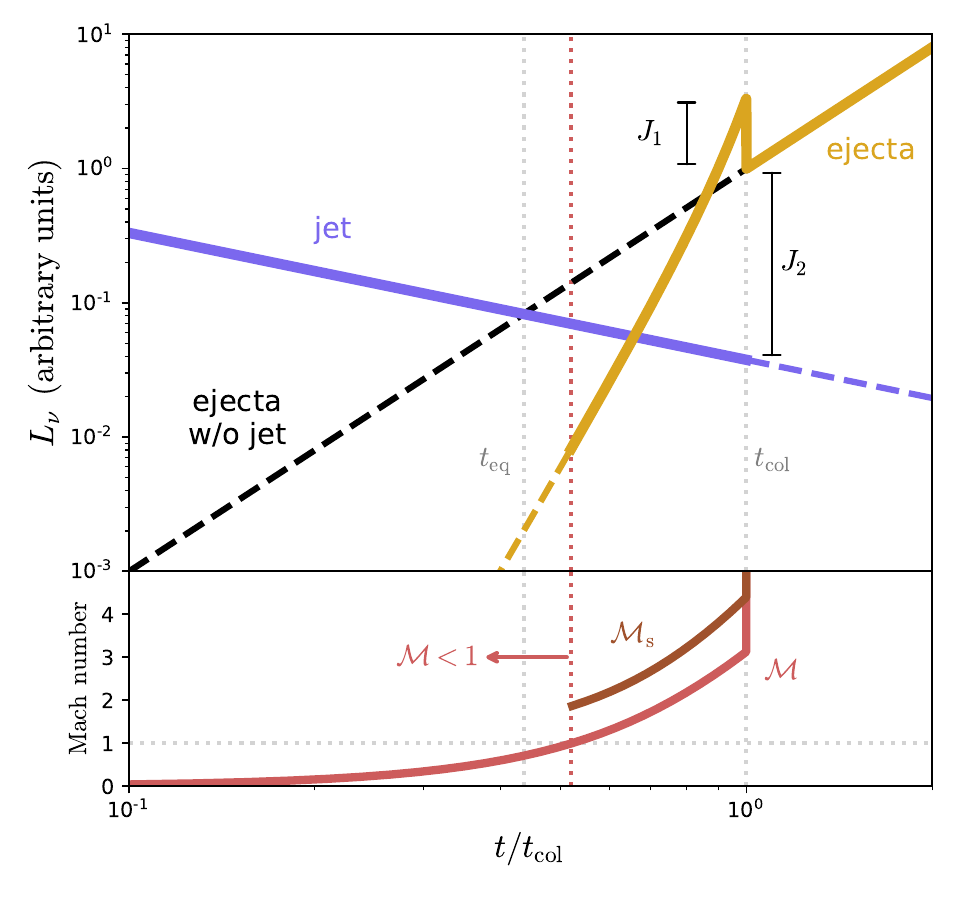}
    \caption{
    {\it Top panel}: synchrotron light-curve resulting from an expanding single-velocity shell (ejecta; yellow) with constant velocity running into an ambient medium shaped by a preceding Sedov-Taylor blast-wave (jet; blue). Both are assumed to be in the deep-Newtonian regime with $p=2.15$. The dashed black curve shows the naive expectation if the ejecta were to run into an unperturbed constant-density ISM instead. The total (combined) light-curve is modified from this naive scenario at times $t_{\rm eq} < t \leq t_{\rm col}$. 
    {\it Bottom panel}: Mach number of the ejecta forward-shock as a function of time. The Mach number is modestly low because the upstream medium is pre-shocked to high temperatures by the jet. At times $t < 0.52 t_{\rm col}$ the Mach number is $\mathcal{M} < 1$ (dotted-red vertical curve) precluding any synchrotron emission from the ejecta at these times.
    }
    \label{fig:numerical_results}
\end{figure}

Equation~(\ref{eq:Lnu_ej_tscaling}) shows that the ejecta radio flare would rise as $t^9-t^{10}$ at $t \ll t_{\rm col}$
so that emission is strongly inhibited at these times.\footnote{We show later in this section that this emission is in fact completely quenched at $t \ll t_{\rm col}$ because the ejecta propagates sub-sonically at these times.}
The top panel of Figure~\ref{fig:numerical_results} shows the resulting light-curve calculated for an ejecta coasting at constant velocity $v_{\rm ej}$ (appropriate in the regime $E_{\rm ej} \gg E_{\rm j}$) within an ambient medium described by the Sedov-Taylor profile (yellow curve). The dashed black curve shows for comparison the afterglow light-curve predicted for the same ejecta expanding into an unperturbed constant-density ISM, as assumed by previous models.

As discussed above and illustrated in Fig.~\ref{fig:numerical_results}, the ejecta light-curve rises steeply near $t \lesssim t_{\rm col}$ and is severely inhibited at earlier times in comparison to unshocked ISM models. However, immediately before $t=t_{\rm col}$ this light-curve is in fact enhanced compared to such models, and compared to the light-curve shortly after $t_{\rm col}$ (which converge at $t>t_{\rm col}$ when the ejecta forward-shock has overtaken the jet and therefore shocks an unperturbed ISM).
This enhancement is due to the larger density of the thin-shell of swept-up ISM immediately behind the jet forward shock $\simeq 4 n$, however, is also partly compensated by the gas bulk velocity behind the jet forward-shock which implies $v_{\rm ej}-v(R_{\rm j}) = 7 v_{\rm ej} / 10$.
From eq.~(\ref{eq:Lnu_scaling}; see also~\ref{eq:Lnu_ej_tscaling}) we find that these two effects induce a jump in the light-curve immediately following $t=t_{\rm col}$ by a factor of
\begin{equation}
\label{eq:J1}
    J_1 \equiv
    \frac{L_{\nu,{\rm ej}}(t_{\rm col}^-)}{L_{\nu,{\rm ej}}(t_{\rm col}^+)} 
    =
    4^\frac{p+5}{4}
    \begin{cases}
    \left(\frac{7}{10}\right)^\frac{5p-3}{2}; &t_{\rm col}<t_{\rm DN}
    \\
    \left(\frac{7}{10}\right)^\frac{p+5}{2}; &t_{\rm col}>t_{\rm DN}
    \end{cases}
    .
\end{equation}
Numerically this results in $J_1 \sim 3$ for values of $p$ close to $p \approx 2$.
The above assumes identical microphysical parameters ($\epsilon_B$, $\epsilon_e$, $p$) for the two shocks which, however, need not be the case. If $\epsilon_B$ and/or $\epsilon_e$ are higher for the ejecta--hot-ISM shock than for the ejecta--unperturbed-ISM shock at $t>t_{\rm col}$ then the jump in luminosity at this transition point would be even larger.

At the same time ($t=t_{\rm col}$) the jet synchrotron afterglow luminosity is one to two orders of magnitude weaker than that of the ejecta. This is easy to estimate since at $t=t_{\rm col}$ the jet and ejecta forward shocks are at the same radius (by definition; eq.~\ref{eq:tcol}), however their velocities differ by a factor $v_{\rm j}/v_{\rm ej} = 2/5$ (eq.~\ref{eq:Rjvj}).
The relative synchrotron luminosity of these components as implied by eq.~(\ref{eq:Lnu_scaling}), is therefore
\begin{equation}
\label{eq:J2}
    J_2 \equiv
    \frac{L_{\nu,{\rm ej}}(t_{\rm col}^+)}{L_{\nu,{\rm j}}(t_{\rm col})} =
    \begin{cases}{}
    \left(\frac{2}{5}\right)^{-\frac{5p-3}{2}}; &t_{\rm col}<t_{\rm DN}
    \\
    \left(\frac{2}{5}\right)^{-\frac{p+5}{2}}; &t_{\rm col}>t_{\rm DN}
    \end{cases}
    .
\end{equation}
For $p=2$ this results in $J_2 \approx 25$ for both regimes (deep-Newtonian and otherwise). For larger values of $p$ this ratio can be significantly larger ($J_2 \approx 40,240$ for $p=3$ in each regime, respectively).

An additional concern for synchrotron emission from the ejecta in this medium is the high temperature of gas previously shocked by the jet.
The sound speed of shocked gas immediately behind the jet-ISM forward shock is given by the Sedov-Taylor solution as
$c_s(R_{\rm j}) = 2 \sqrt{2g(g-1)} R_{\rm j} / 5 (g+1) t$.
Gas immediately behind the jet-ISM shock front has a velocity $v(R_{\rm j}) = 4 R_{\rm j} / 5 (g+1) t$.
The Mach number of the ejecta with respect to the thin shell of swept-up ISM immediately preceding the jet forward-shock is therefore 
\begin{equation}
\label{eq:Mach}
    \mathcal{M}(R_{\rm j}) = \frac{v_{\rm ej}-v(R_{\rm j})}{c_s(R_{\rm j})} 
    = \frac{7(g+1)}{4\sqrt{2g(g-1)}}
    \approx 3.13 
    ,
\end{equation}{}
and decreases at $r<R_{\rm j}$. In the last equality we have assumed $g=5/3.$
Numerically, we find that $\mathcal{M} \leq 1$ for $t \leq 0.52 t_{\rm col}$ (see Fig.~\ref{fig:numerical_results}). Thus, at early times a shock does not form and non-thermal synchrotron emission would not be produced by the ejecta.

We note that in the literature a different Mach number $\mathcal{M}_{\rm s}$ is often defined, as the ratio of upstream bulk velocity to upstream sound speed in the rest frame of the shock-front (instead of as defined above for $\mathcal{M}$ --- in the downstream rest frame). The two are simply related using the shock compression ratio $r(\mathcal{M}_{\rm s}) = (g+1)/(g-1+2/\mathcal{M}_{\rm s}^2)$ via the implicit relation $\mathcal{M}_{\rm s} = \mathcal{M} r/(r-1)$. In the bottom panel of Fig.~\ref{fig:numerical_results} we also plot $\mathcal{M}_{\rm s}$ (solid yellow). Under this definition, the Mach number at $t=t_{\rm col}$ is $\mathcal{M}_{\rm s}(R_{\rm j}) \approx 4.40$.


From a theoretical standpoint, low Mach-number shocks face problems producing a non-thermal population of relativistic electrons necessary for synchrotron emission. The reason has to do with the fact that standard diffusive shock (first-order Fermi) acceleration requires electrons reflect off MHD waves in the upstream/downstream medium, however --- if the electron gyro-radius is $\ll$ than the shock-thickness ($\sim$ ion gyro-radius) then such diffusive reflection cannot occur. For sufficiently strong shocks the thermal pool of shock-heated electrons extends to $\gamma \gg 1$, and these electrons can participate in diffusive shock acceleration. For weak shocks however, this is not the case, leading to the so-called `injection problem', namely whether and how electrons are pre-accelerated to $\gamma \gg 1$ where they can then undergo standard diffusive shock acceleration.
\cite{Guo+14a,Guo+14b} showed using multi-dimensional particle-in-cell simulations that particle acceleration can occur even for low Mach-number shocks and identified the shock drift acceleration mechanism operating to pre-accelerate electrons and overcome the injection problem.
However, recent work suggests that this mechanism may only be effective in shocks with $\mathcal{M}_{\rm s} \gtrsim 2.3$ \citep{Kang+19}. In our scenario, this would inhibit emission by the ejecta at times $t \lesssim 0.62 t_{\rm col}$ when $\mathcal{M}_{\rm s} \lesssim 2.3$ (see fig.~\ref{fig:numerical_results}).

Observationally, low Mach-number shocks in galaxy clusters (even those with inferred $\mathcal{M}_{\rm s}<2.3$) are bright non-thermal synchrotron sources, implying that electron acceleration in such settings can occur in Nature.
Still, there are indications that this emission is in tension with predictions of diffusive shock acceleration and may therefore require an alternative model \citep[e.g.][]{Vazza&Bruggen14,Botteon+20}. One possibility is the acceleration of relic $\gamma \gg 1$ `fossil' electrons that inhabit the upstream medium, which however requires an explanation as to the origin of these fossil electrons.
This would be a natural consequence in our scenario since a non-thermal population of such relic $\gamma \gg 1$ electrons is produced by the jet-ISM forward shock. This could potentially alleviate issues associated with the injection problem and allow for efficient synchrotron emission from the ejecta forward shock for even very low Mach numbers (as plotted in fig.~\ref{fig:numerical_results}), but this is still an open research area.

Regardless of whether particle acceleration is effective at low ($\lesssim 2.3$) Mach numbers it seems plausible that near $t=t_{\rm col}$ when $\mathcal{M}_{\rm s} \approx 4$, synchrotron emission from the ejecta forward shock would be produced as estimated in our above analysis. Our qualitative results therefore remain unchanged --- at times $t \ll t_{\rm col}$ synchrotron emission by the ejecta is significantly inhibited (or even completely shut-off) compared to the expected emission if the ejecta were to expand in an unperturbed ISM, while immediately before $t=t_{\rm col}$ the synchrotron light-curve should rise steeply and reach a local maximum that is a factor $\sim 3$ higher than immediately after $t=t_{\rm col}$ (eq.~\ref{eq:J1}).

Finally, we note that diffusive-shock acceleration predicts that the power-law index $p$ of accelerated electrons depend on the shock Mach number. In the test-particle limit, and for non-relativistic shocks --- diffusive-shock acceleration predicts \citep[e.g.][]{Blandford&Eichler87} that
$p = 2 (\mathcal{M}^2+1)/(\mathcal{M}^2-1)$. Here we have implicitly assumed an adiabatic index $g=5/3$. In the strong shock limit ($\mathcal{M} \to \infty$) this expression reduces to the familiar result $p=2$, and the particle spectrum steepens with decreasing Mach number (see \citealt{Steinberg&Metzger20} for a related discussion in the context of classical novae). On both observational and theoretical grounds, the naive $p=2$ prediction of this theory is likely a lower-limit on the true value of $p$, and radio SNe exhibit a diversity in spectral indices with $p>2$ (Type Ib/c radio SNe seem to favor $p \approx 3$, whereas Type II radio SNe are better fit with lower values of $p$; \citealt{Weiler+02}). Nevertheless, we might expect that the qualitative trend of increasing $p$ with lower $\mathcal{M}$ remain valid in the full non-linear regime.
In our scenario, steeper electron spectra at $t < t_{\rm col}$ would act to further inhibit the synchrotron emission at observed frequencies $\nu \gg \nu_m$. The monotonic decrease in $p(t)$ towards $t=t_{\rm col}$ (as $\mathcal{M}$ increases) would cause the light-curve at a fixed band to increase even more sharply as a function of time than our constant-$p$ estimate shown in Fig.~\ref{fig:numerical_results}. A rapidly softening spectrum would be another interesting and unique feature that we therefore posit could accompany the locally-peaking ejecta radio-flare at $t=t_{\rm col}$.

\section{Implications}
\label{sec:Implications}

In the following we apply our results to astrophysical sources and discuss their observational implications.

\subsection{GW170817}

The interaction of the slow kilonova ejecta with an ISM that is pre-shocked by the successful jet seems both natural and unavoidable for events like GW170817. In the previous sections we have shown that this has the effect of quenching the ejecta afterglow signature at early times when the ejecta is still expanding within the cavity generated by the jet-ISM forward shock.

For GW170817, the jet energy and ambient ISM density are constrained by radio and X-ray observations of the jet afterglow \citep{Margutti+17,Troja+17,Alexander+17,Haggard+17,Alexander+18,Margutti+18,Dobie+18,D'Avanzo+18,Troja+18,Mooley+18a,Mooley+18b,Mooley+18c,Granot+18,Fong+19,Ghirlanda+19,Troja+19,Hajela+19}. The two parameters are degenerate with one another to a large extent and have significant uncertainties from the modelling, however their ratio is somewhat better constrained. 
For example, \cite{Mooley+18b} find $E_{\rm j}/n \sim 10^{53} \, {\rm erg \, cm}^3$.
The kilonova ejecta bulk-velocity is inferred from the optical--near-infrared kilonova observations to be\footnote{High resolution numerical simulations \citep[e.g.][]{Kiuchi+2017} reveal a fast tail moving at  higher velocity (up to $\sim0.8c$). However, the amount of matter and energy in this fast tail  is negligible and irrelevant for our late time considerations.} $v_{\rm ej} \approx 0.1-0.3 c$ \citep[e.g.][]{Villar+17}.
From eq.~(\ref{eq:tcol}) it follows that the collision time between the ejecta and jet forward-shock should be
\begin{equation}
\label{eq:tcol_GW170817}
    t_{\rm col} \approx 80 \, {\rm yr} \, \left(\frac{v_{\rm ej}}{0.2 c}\right)^{-5/3}
    \left(\frac{E_{\rm j}/n}{10^{53}\,{\rm erg \, cm}^3}\right)^{1/3} .
\end{equation}

The expectation is therefore that the X-ray and radio afterglow signal of GW170817 will continue to decline until $\sim 80 \, {\rm yr}$ post merger (with large uncertainties), after which a sharp rise 
in the afterglow luminosity, by a factor of $J_1 \times J_2 \sim 90$ (eqs.\ref{eq:J1},~\ref{eq:J2}), is expected as the merger ejecta overtakes the jet forward shock.
In the idealized framework discussed above, measurement of $t_{\rm col}$ would provide a strong constraint on $v_{\rm ej}$ (eq.~\ref{eq:tcol_GW170817}).

Using the results of \cite{Sironi&Giannios13} (c.f. their eq.~12) with the best fit parameters for the GW170817 jet afterglow from \cite{Mooley+18b}, we find that the jet component's emission is 
$F_{\nu,{\rm j}}^{\rm GHz}(t=80\,{\rm yr}) \approx 0.3 \mu {\rm Jy}$ around $t_{\rm col}$. 
This is well below the detection threshold of the most sensitive current-day radio telescopes, however the ejecta radio flare is expected to reach a much higher peak luminosity of $\sim 25 
\mu {\rm Jy}$ at $t=t_{\rm col}$ (see eq.~\ref{eq:Fnu_tcol}). This would be easily detectable with next-generation radio facilities such as SKA or ngVLA which would be online well before $\sim$2090 (when $t=t_{\rm col}$ for GW170817).

The kilonova ejecta radio flare will peak on its Sedov-Taylor deceleration timescale, which is of order $\sim 100$ yrs for the low ISM density inferred around GW170817 (eq.~\ref{eq:tdec}). On shorter timescales more relevant for near-future detection, the rising ejecta signature would naively be expected to pop up above the declining jet-afterglow after $t > t_{\rm eq} \approx 0.44 t_{\rm col} \approx 35 \, {\rm yr} \, (v_{\rm ej}/0.2c)^{-5/3}$ (eqs.~\ref{eq:teq},\ref{eq:tcol_GW170817}). Instead, we have argued in this work that the ejecta signature would be inhibited on these timescales and thus we would not expect the ejecta radio flare to show up so early. 
This is particularly relevant for efforts at constraining the kilonova ejecta properties using non-detections of a rising ejecta radio flare \citep[e.g.][]{Kathirgamaraju+19,Hajela+19}.
A lacking detection of a rise in the radio light-curve before $t_{\rm col} \sim 80$ \, {\rm yr} is in fact expected based on our current (albeit idealized) analysis, and therefore does not necessarily constrain the ejecta parameters.

An important caveat is our assumption of spherical symmetry. For GW170817, the time at which the jet becomes non-relativistic can be estimated from eq.~(\ref{eq:tsph}) to be $t_{\rm NR} \approx 30 \, {\rm yr}$, where we have taken $\theta_{\rm j} = 0.04$ as a fiducial value based on the best-fitting model of \cite{Mooley+18b}
. The jet completes its azimuthal expansion around the same time to within a factor of a few, $t_{\rm sph} \gtrsim t_{\rm NR}$. This can become comparable to $t_{\rm eq}$ and so care is needed in interpreting the results.
Secondly, we have discussed idealized single-velocity component ejecta whereas realistic kilonova ejecta  should have both radial and azimuthal density stratification \citep[e.g][]{Radice+18,Gottlieb+18}. Furthermore, constraining the synchrotron emission from a fast tail of this ejecta would require extending our results to trans-relativistic regimes \citep[e.g.][]{Hotokezaka&Piran15,Hotokezaka+18,Kathirgamaraju+19}. This is straightforward for a spherically symmetric model, however the early timescales on which emission from such trans-relativistic material would be relevant are almost certainly $< t_{\rm sph}$ so that the spherical assumption for relativistic ejecta does not make much sense.
For these reasons, we plan to extend our current analysis in future work investigating the multi-dimensional nature of this problem.

\subsection{Future BNS mergers}

As discussed in the previous subsection, the ejecta radio flare and late-time jet afterglow emission from GW170817 is expected to be relatively weak and evolve over long timescales. This is a direct consequence of the low ISM density at the location of the merger, $n < 10^{-2} \, {\rm cm}^{-3}$ \citep{Hajela+19}.
If a future BNS merger occurs in a denser environment, then the ejecta radio flare would be brighter ($L_\nu \propto n^{(3+3p)/20}$; \citealt{Sironi&Giannios13}) and evolve on shorter timescales ($t_{\rm col} \propto n^{-1/3}$). As an illustrative example, if an event identical to GW170817 
but at a distance of $120 \, {\rm Mpc}$ (characteristic of the LIGO O3 horizon distance; \citealt{LVC18})
were to occur instead in an environment with $n = 1 \, {\rm cm}^{-3}$, we would predict $t_{\rm col} \approx 8 \, {\rm yr} \, (v_{\rm ej}/0.2c)^{-5/3}$ and a (local) peak flux in the deep-Newtonian regime,\footnote{The prefactor can differ by a factor of a few for different values of $p$.} 
\begin{align}
\label{eq:Fnu_tcol}
    F_{\nu,{\rm ej}}\left(t_{\rm col}\right) 
    \approx 0.6 \, {\rm mJy} \,
    &n_0^\frac{p+1}{4} E_{{\rm j},50} \left(\frac{v_{\rm ej}}{0.2c}\right)^\frac{p+1}{2}
    \\ \nonumber
    &\times \bar{\epsilon}_{e,-1} \epsilon_{B,-2}^\frac{p+1}{4} \nu_{\rm GHz}^{-\frac{3(p+1)}{10}} \left(\frac{d}{120 \, {\rm Mpc}}\right)^{-2} 
    .
\end{align}
This is easily observable, and shows that BNS mergers detected in the near future may be even more promising sources than GW170817.

\subsection{(on-axis) SGRB radio follow-up}


In recent years, late-time radio follow-up of nearby short GRBs (SGRBs) has been conducted by several authors  \citep{Metzger&Bower14,Horesh+16,Fong+16,Klose+19}. These observations have all resulted (so far) in non-detections, that have been used to place constraints on combinations of the mass and energy of a possible associated kilonova ejecta, the ambient ISM density, and microphysical parameters. These observations are motivated by a class of theoretical models which assert that (possibly a subset of-) SGRBs are produced by a highly-magnetized rapidly-rotating NS (the `magnetar model'; \citealt{Metzger+11,Rowlinson+13}). In this scenario it is expected that the magnetar's rotational energy will, at least in part, be deposited into its surroundings and increase the kilonova ejecta's energy by an order of magnitude or more. This energy boost would enhance the ejecta radio-flare significantly, and thus current upper-limits on such emission manage to stringently constrain this scenario.

Our present work could imply that the expected kilonova radio-flare may be inhibited due to the jet pre-shocking of the ISM, which might naively allow for higher $E_{\rm ej}$ to still be consistent with the radio non-detections and limit the constraints on the magnetar model.
However, there is only a narrow parameter space where this might apply, given that the large $E_{\rm ej}$ correspond to a highly-accelerated ejecta with $\beta_{\rm ej} \sim 1$. This makes the relevant timescales $\sim t_{\rm col}$ very short (eq.~\ref{eq:tcol}), so that current observations may already be past $t>t_{\rm col}$ when the ejecta radio-flare light-curve is no more affected by the jet. 
Furthermore, our assumption of spherical symmetry, which is already marginal for typical BNS merger parameters, certainly breaks down for such fast kilonova ejecta velocities (see Fig.~\ref{fig:spherical_assumption} and discussion in \S\ref{sec:Discussion}). In such systems the jet is viewed $\sim$on-axis given that these were detected as classical SGRBs. In this sense, along the line of sight, the jet already obscures the ejecta and there is no need to wait till $t \gtrsim t_{\rm sph}$ for this to occur. However, regions of the ejecta at other angles will expand relatively unobstructed into the ISM. Since emission from these ejecta components is not beamed (unless, perhaps, for extreme values of $E_{\rm ej}/M_{\rm ej}$) and because the ``obscuring jet'' is optically-thin to synchrotron self-absorption, emission from such ejecta would contribute to the light-curve even at times $t < t_{\rm sph},t_{\rm col}$.

For these reasons, we expect that constraints placed on {\it highly-energetic} kilonova ejecta remain valid and are not severely impacted by the jet-ejecta coupling discussed in this work. In contrast, we caution that constraining ejecta properties for un-boosted (typical) kilonovae ejecta likely does need to account for such effects. Most SGRBs are too distant for current radio upper-limits to significantly constrain such ``standard'' kilonovae ejecta at present, however future radio facilities may be able to do so in the future.

\subsection{LGRBs and broad-lined Type-Ic SNe}
\label{subsec:LGRB_IcSNe}

The scenario described in this paper is also relevant to long GRBs (LGRBs), which are known to be accompanied by very energetic broad-lined Ic SNe \citep{Galama+98,Bloom+99,Hjorth+03,Woosley&Bloom06}.
Motivated by this scenario, \cite{BarniolDuran&Giannios15} examined the radio re-brightening of LGRB afterglows from interaction of the accompanying SN with the ambient ISM. Applying their model to late-time radio data of LGRBs available at the time and using the non-detection of such re-brightening for any of their sources, \cite{BarniolDuran&Giannios15} constrained the ambient density for GRB 030329 and predicted that it's SN radio signature should become detectable by $\sim$2030 at the latest.
Subsequent work by \cite{Kathirgamaraju+16} extended these results to SNe ejecta exhibiting a velocity profile, and extending into the relativistic regime. These results were then used in \cite{Peters+19}, who presented deep new upper-limits on radio emission from LGRBs, to place significant constraints on these source's properties.

In the following we briefly revisit these constraints in light of our current results. Since our expectation is that any SN-ejecta radio-flare would be inhibited at times $<t_{\rm col}$ compared to the predictions of previous models, the inferred constraints from radio non-detections may be overly stringent.

For GRB 030329, inferred parameters of the jet indicate $E_{\rm j}/n \sim (0.8-5) \times 10^{51} \, {\rm erg \, cm}^3$ (\citealt{Pihlstrom+07,Mesler+12}; see also \citealt{Mesler&Pihlstrom13}). Observations of the accompanying SN 2003dh constrained the ejecta velocity to $v_{\rm ej} \approx ( 29 \pm 5.8 ) \times 10^3 \, {\rm km \, s}^{-1}$ \citep{Mazzali2007}. From eq.~(\ref{eq:tcol}) we therefore find that $t_{\rm col} \approx 40-140 \, {\rm yr}$, much later than the latest epoch of observation of this source \citep{Peters+19}. We therefore conclude that radio re-brightening by the SN ejecta should not have occurred yet for GRB 030329, and that a non-detection of such re-brightening does not currently constrain the source properties.
Note also that this statement is independent of the assumed microphysical parameters $\epsilon_e, \epsilon_B$ and does not depend on the poorly constrained ambient density, but on the slightly better constrained $E_{\rm j}/n_0$ (this is in contrast to the standard picture of non-interacting jet and SN).

Another relevant point is the fact that, on theoretical grounds, LGRB jets and their associated SNe ejecta are thought to propagate into an ambient stellar-wind environment rather than a constant-density ISM. The situation is complicated because there is no strong observational evidence for a wind-like $\rho \propto r^{-2}$ density profile in LGRB afterglow light-curves. Nonetheless, in Appendix~\ref{sec:Appendix_winddensity} we extend the results of the preceding sections to a wind circum-stellar density profile for completeness.

Finally we note that two important related caveats arise in the context of LGRBs and also for superluminous-SNe, discussed in the next section, if the latter harbor relativistic jets. In both cases the jet deposits significant energy into a cocoon
\citep{Nakar+17}. This generates a highly anisotropic high-velocity ejecta that might complicate the above picture. In fact some evidence for such an outflow with velocities up to $0.2c$ has been observed in several luminous SNe \citep{Piran+19}. Clearly addressing the  evolution of such a system requires a detailed numerical simulation. 

\subsection{Other jetted events}

We have focused on the BNS merger or LGRB-SN scenario as a main motivation for this work, however the ideas discussed above can potentially be applicable to other settings as well.
For example, superluminous-SNe (SLSNe; \citealt{Quimby+11,Gal-yam12}) are extremely energetic SNe whose optical light-curve is thought to be powered by interaction with circum-stellar material \citep{Chevalier&Irwin11,Ginzburg&Balberg12} or by a central engine \citep{Kasen&Bildsten10,Woosley10,Dexter&Kasen13}. Motivated by connections between SLSNe and broad-lined SNe Ic that are accompanied by LGRBs \citep[e.g][]{Metzger+15}, \cite{Margalit+18} proposed a mechanism by which collimated relativistic outflows may be launched in conjunction with SLSNe. Despite significant observational follow-up in X-ray and radio bands \citep{Coppejans+18,Margutti+18_SLSNe,Law+19}, only a single SLSN, PTF10hgi, has been observed as a source of late-time non-thermal emission \citep{Eftekhari+19,Law+19}. The current radio data is sparse, but one possible interpretation is the signal being an off-axis afterglow from a jet associated with this SLSN.\footnote{An alternative interpretation of the signal is plerionic emission from the central engine, and its possible relation to fast radio burst progenitors \citep{Metzger+17,Margalit&Metzger18,Eftekhari+19,Law+19}}
If indeed a subset of SLSNe are accompanied by jets, then the analysis above would be also relevant for the late-time radio flares of such SN ejecta.

These ideas may also be relevant to the interpretation of decade-long radio transients. Recently \cite{Law+18} identified FIRST J141918.9+394036, a long-duration declining radio transient that is consistent with an off-axis `orphan' LGRB. According to the modeling presented in \cite{Law+18}, an initially off-axis GRB with total energy $E_{\rm j} = 10^{51} \, {\rm erg}$ and a surrounding ISM density of $n = 10 \, {\rm cm}^{-3}$ can fit the data for an initial explosion epoch around $\sim$1993. This interpretation has been provided additional supporting evidence by observations with the European VLBI Network (EVN), which resolve the source size, finding $R_{\rm j} = 1.6 \pm 0.3 \, {\rm pc}$ \citep{Marcote+19}.
In this scenario, it would generally be expected that an energetic GRB-SN accompany the event. Thus, at late enough epochs the SN radio flare should outshine the GRB afterglow (eq.~\ref{eq:teq}) and produce a rising light-curve \citep{BarniolDuran&Giannios15}. The non-detection of such a rise can be accommodated for a more prolonged period of time given our current analysis (up till $t \lesssim t_{\rm col}$ rather than $t<t_{\rm eq}$), and this allows us to place constraints on the properties of a putative SN ejecta associated with J1419. The source has an initial detection epoch in 1993, which provides a lower-limit on the source age $t_{\rm min}$. At the time of the EVN observation, this corresponds to $t_{\rm min} \approx 25 \, {\rm yr}$. Demanding that $t_{\rm min} < t_{\rm col}$, or equivalently that $v_{\rm j} t_{\rm min} < R_{\rm j}$ we find that $v_{\rm ej} < 62,600 \, {\rm km \, s}^{-1}$. This is somewhat larger than typical broad-line Ic SNe velocities $\sim 20,000 \, {\rm km \, s}^{-1}$ \citep{Modjaz+16}, and therefore currently consistent with the scenario. 

Given it's nearby distance ($87 \, {\rm Mpc}$) and age, J1419 would be an ideal target for detecting the SN radio-flare re-brightening in the future.
In particular, using the EVN measurement of $R_{\rm j}$ at time $t_{\rm EVN}$ along with eqs.~(\ref{eq:Rjvj},\ref{eq:tcol}), we predict a re-brightening around $ \lesssim t_{\rm col} \approx 86^{+28}_{-25} \, {\rm yr} \, (t_{\rm EVN}/ 25 \, {\rm yr})^{-2/3} \beta_{{\rm ej},-1}^{-5/3}$. This could be as early as $\sim$2050, or possibly sooner if the ejecta velocity is higher or the explosion epoch precedes 1993 (it cannot be later because $t_{\rm EVN} \geq t_{\rm min} \approx 25 \, {\rm yr}$). The rising ejecta-flare signature would be easily detectable with even low-sensitivity radio facilities (we roughly estimate the $1 \, {\rm GHz}$ flux at this time to be $\sim 20 \, {\rm mJy}$; eq.~\ref{eq:Fnu_tcol}), and we therefore encourage continuous follow-up of J1419 to constrain this scenario.

\section{Discussion}
\label{sec:Discussion}

In this paper we have argued that predictions for radio-flares produced by kilonovae or GRB-SNe ejecta must be modified at early (but not too early) times $t_{\rm sph} < t \lesssim t_{\rm col}$ (eq.~\ref{eq:tcol}) because the medium into which such ejecta expand is pre-shaped by the jet that accompanies such events. In particular, we have shown that the radio-flare of such ejecta would be: significantly inhibited at $t \ll t_{\rm col}$; revert back to standard predictions at $t \gtrsim t_{\rm col}$; and possibly experience a sharp local peak and enhancement at $t \lesssim t_{\rm col}$ (Fig.~\ref{fig:numerical_results}).
In our present analysis we have, for simplicity, assumed spherical symmetry, however there are clear caveats to this approach.

\begin{figure}
    \centering
    \includegraphics[width=0.45\textwidth]{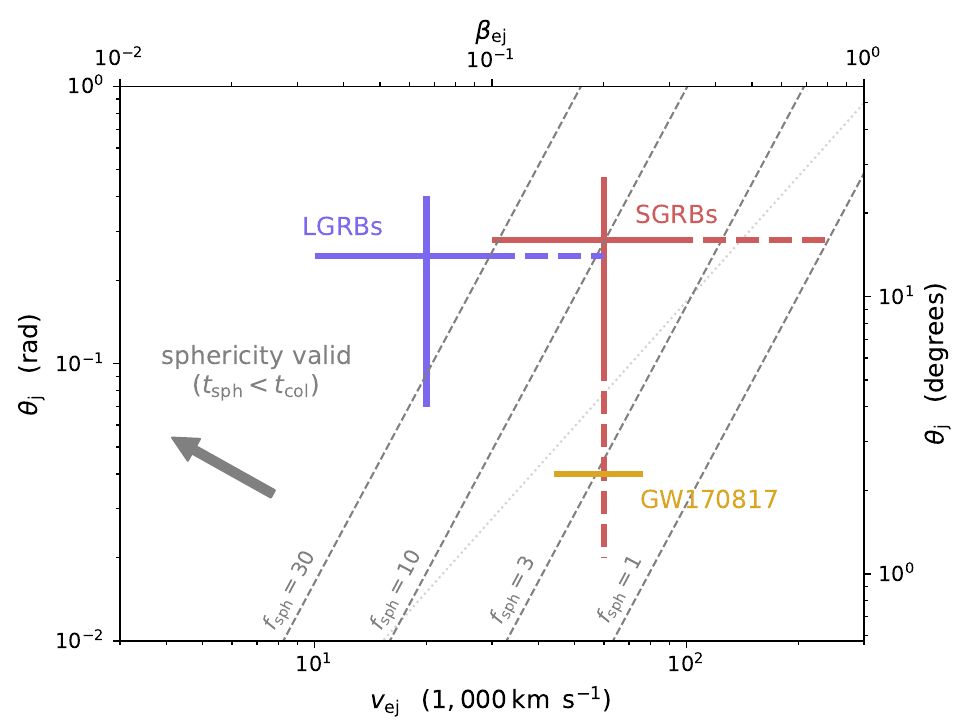}
    \caption{Plane of initial jet opening angle $\theta_{\rm j}$ and ejecta velocity $v_{\rm ej}$ showing characteristic values for LGRBs (blue), SGRBs (red), and GW170817 (yellow). The spherically-symmetric approach to the problem is valid to the left of a dashed line (eq.~\ref{eq:sph_assumption}), depending on the somewhat uncertain value of $f_{\rm sph} \equiv t_{\rm sph}/t_{\rm NR} \gtrsim 1$. Narrowly collimated jets and/or fast ejecta do not obey these requirements, motivating future multi-dimensional numerical work extending our current analysis. The light-grey dotted curve shows the analogous condition for a wind-like ambient density profile that may be relevant for LGRBs, and taking $f_{\rm sph}=10$ (see Appendix~\ref{sec:Appendix_winddensity}).}
    \label{fig:spherical_assumption}
\end{figure}

The spherically-symmetric assumption is relevant at times $t \gtrsim t_{\rm sph}$ after the initially-collimated jet decelerates and azimuthally expands into a quasi-spherical configuration (eq.~\ref{eq:tsph}). This assumption is valid at the characteristic timescale $t_{\rm col}$ if $t_{\rm sph} < t_{\rm col}$. This depends only on the initial jet opening angle $\theta_{\rm j}$ and the ejecta velocity $v_{\rm ej}$,
\begin{equation}
\label{eq:sph_assumption}
    \theta_{\rm j} \gtrsim 
    0.049 \, {\rm rad}
    \, \left(\frac{f_{\rm sph}}{10}\right)^{3/2} \beta_{{\rm ej},-1}^{5/2} ,
\end{equation}
where $f_{\rm sph} \sim 3-10$ \citep{DeColle+12,Granot&Piran12,Duffell&Laskar18} is a numerical factor relating the non-relativistic and 
spherical
timescales of the jet, $t_{\rm sph} = f_{\rm sph} t_{\rm NR}$ (eq.~\ref{eq:tsph}).
Equation~(\ref{eq:sph_assumption}) is a strict lower limit imposing the minimal requirement --- that spherical symmetry reasonably describe the system at $t=t_{\rm col}$. Requiring that the spherical assumption be valid at earlier times $t<t_{\rm col}$ would imply a  larger  minimal opening angle by a factor $(t/t_{\rm col})^{-3/2}$. Importantly, imposing this condition at time $t=t_{\rm eq}$ (eq.~\ref{eq:teq}), when the light-curve we predict first deviates from earlier models (Fig.~\ref{fig:numerical_results}), demands $\theta_{\rm j} \gtrsim 0.16 \, {\rm rad} \, (f_{\rm sph}/10)^{3/2} \beta_{{\rm ej},-1}^{5/2}$ (with some dependence on $p$ that has been omitted here for brevity).

Figure~\ref{fig:spherical_assumption} shows the parameter-space of initial jet opening-angle and ejecta velocity for LGRBs and their associated SNe \citep{Fong+15,Modjaz+16}, cosmological SGRBs and their associated kilonovae ejecta \citep{Fong+15}, and for GW170817 \citep{Mooley+18b,Villar+17}.
These estimates have considerable uncertainty, and in particular there are suggestions of fast-tail ejecta for both SNe \citep{Piran+19} and kilonovae ejecta \citep{Kiuchi+2017} illustrated as dashed extensions of the errorbars. Dashed grey curves show the minimal jet opening-angle such that spherical symmetry is applicable at $t=t_{\rm col}$ (eq.~\ref{eq:sph_assumption}) for different values of $f_{\rm sph}$.
For fast ejecta and/or very narrowly-collimated jets this condition is not satisfied and multi-dimensional numerical tools must be used to investigate the joint jet-afterglow ejecta-radio-flare light-curves. This will be studied in upcoming future work that will extend our current analytic treatment.
In particular, the spherical condition is not satisfied for GW170817 canonical parameters unless $f_{\rm sph} \lesssim 3$. For LGRBs on the other hand, the spherically-symmetric model discussed in the present work may be more readily applicable.
Qualitatively, for systems where the jet has not yet fully sphericized we expect that the true ejecta radio flare be inhibited at early times by a factor $\sim \Omega/4\pi$, where $\Omega$ is the solid-angle subtended by the jet at the time of interest. Thus the true result in such scenarios should lie in between our current estimates and those neglecting jet-ejecta interaction.

Multi-dimensional numerical work into this problem will also allow investigation of more realistic ejecta, characterized by a velocity profile rather than a single `bulk' velocity (see Appendix~\ref{sec:Appendix_velocityprofile}). We also posit that additional features in the radio light-curve may be produced when portions of the azimuthally-expanding jet forward-shock collide with the ejecta and/or the opposing counter-jet. Such features may provide additional diagnostics of the system properties but can only be probed using multi-dimensional numerical simulations.

The efficiency of particle acceleration in our scenario is another interesting question that we encourage future work to examine in greater detail. The shock formed by the ejecta colliding with the hot pre-shocked medium is characterized by a low Mach number (eq.~\ref{eq:Mach}; Fig.~\ref{fig:numerical_results}). The presumably-enhanced magnetic turbulence in this pre-shocked medium may also play a role in changing the characteristics and efficiency of particle acceleration in this `dual-shock' scenario (as can the presence of `fossil' high-energy electrons accelerated by the preceding jet-ISM forward shock). An analogous scenario has been discussed in the Solar-physics community where one model for Ground Level Enhancement events (events in which the flux of Solar Energetic Particles is greatly enhanced) posits that shocks between consecutive colliding coronal mass ejections can more efficiently accelerate particles \citep{Li+12,Zhao&Li14,Wang+19}.
In our context, the efficiency of $e^-$-acceleration and effective $\epsilon_e$, $\epsilon_B$, and $p$, at such shocks would influence the prominence and detectability of the predicted peak in the light-curve at $t = t_{\rm col}$ (see Fig.~\ref{fig:numerical_results} and eq.~\ref{eq:J1}).
If conditions in such `dual-shocks' are indeed conducive to particularly effective particle acceleration, this may also have interesting implications for the astrophysical sites of cosmic-ray acceleration.

Our prediction of a sharply-rising and abruptly-declining local peak in the light-curve when the ejecta catches-up with the thin-shell of ISM swept-up by the jet (at $t=t_{\rm col}$) provides motivation for observing strategies that differ from standard logarithmically-spaced intervals. This feature in the light-curve might be missed if the observational cadence is too low. 
Note however that some of the sharp features in the light curve illustrated in Fig.~\ref{fig:numerical_results} should be smoothed out given light travel-time effects which limit $\Delta t \gtrsim R/c \approx \beta_{\rm ej}$. Because $\beta_{\rm ej} \sim 0.1$ or larger in the astrophysical settings we have considered, the light-curve cannot vary on timescales faster than tens of per-cent of the system age.\footnote{A velocity profile and/or some level of asphericity in the ejecta would likely also contribute to smoothing out these sharp features.}
The timing and prominence of this peak can potentially shed light on important properties such as the ejecta velocity (eq.~\ref{eq:tcol}). 
If the local peak at $t=t_{\rm col}$ can be compared to the subsequent global peak of the ejecta radio-flare at $t=t_{\rm dec}$, one can infer the ratio of jet and ejecta energies, 
$E_{\rm j}/E_{\rm ej} \approx ( 0.5 t_{\rm col}/t_{\rm dec} )^3$
(eqs.~\ref{eq:tcol},\ref{eq:tdec}). A strength of this method is the fact that it does not depend on the ambient ISM density or on uncertain microphysical parameters.
For GW170817 the inferred ambient density is particularly low, so that this peak is expected to be faint and occur only many decades from now (eq.~\ref{eq:tcol_GW170817}). However, future BNS mergers may occur in regions with larger ambient densities, improving these prospects (eq.~\ref{eq:Fnu_tcol}).

Finally, we note that \cite{Coughlin19} recently derived relativistic corrections to the self-similar \cite{Blandford&McKee76} solution relevant for shocks with velocities $\sim$several$\times 0.1c$. In our current analysis we used the standard Blandford-McKee solution for the jet forward shock at times $t \gtrsim t_{\rm sph}$, however in the context of BNS mergers the jet forward-shock velocity at times of relevance ($\sim t_{\rm col}$) can be well within the regime affected by such relativistic corrections. Qualitatively, these corrections: reduce the post-shock density at $r \ll R_{\rm j}$ and enhance it near $r = R_{\rm j}$; reduce the post-shock fluid velocity; reduce the post-shock pressure (/temperature). Thus, interestingly, all three effects would contribute towards strengthening the peak luminosity when the ejecta runs into the thin shell of ISM swept up by the jet (eq.~\ref{eq:Fnu_tcol}; this would be due to an increase in $J_1$, c.f. eq.~\ref{eq:J1}).

\section*{Acknowledgements}

BM thanks Eliot Quataert, Brian Metzger, Lorenzo Sironi, Aaron Tran, Stephen Ro, Adithan Kathirgamaraju and Paz Beniamini for helpful conversations and comments.
This work was conceived in interactions that were funded by the Gordon and Betty Moore Foundation through Grant GBMF5076.
This research was supported in part by 
NASA through the NASA Hubble Fellowship grant \#HST-HF2-51412.001-A awarded by the Space Telescope Science Institute, which is operated by the Association of Universities for Research in Astronomy, Inc., for NASA, under contract NAS5-26555 (BM),
and by an Advanced ERC grant TReX (TP).




\bibliographystyle{mnras}
\bibliography{bib} 




\appendix

\section{Ejecta Velocity Distribution}
\label{sec:Appendix_velocityprofile}

In the following we extend our discussion from an idealized `single-velocity shell' ejecta, to one characterized by a velocity distribution. This can be conveniently expressed by the cumulative kinetic energy of ejecta with velocity greater than $v$, $E_{\rm ej}(\geq v)$.
For ease and analytic tractability we focus on a power-law distribution
\begin{equation}
\label{eq:Appendix_Egeqv}
    E_{\rm ej}(\geq v) = E_{\rm ej} \left(\frac{v}{v_{\rm ej}}\right)^{-\alpha}
\end{equation}{}
for $v>v_{\rm ej}$ and $E_{\rm ej}(\geq v) = E_{\rm ej}$ otherwise. We neglect relativistic effects and assume the velocities of relevance are at most trans-relativistic.

The dynamics of the forward shock between the ejecta and a cold constant density external medium of number density $n$ are governed by the differential equation \citep{Piran+13}
\begin{equation}
    \frac{4\pi}{3} n m_p R(t)^3 v(t)^2 \sim E_{\rm ej}\left[\geq v(t)\right] ,
\end{equation}{}
which, for a distribution given by equation~(\ref{eq:Appendix_Egeqv}) results in
\begin{equation}
\label{eq:Appendix_Rv}
    R \propto t^\frac{\alpha+2}{\alpha+5}; ~~~~~
    v = \frac{\alpha+2}{\alpha+5} \frac{R(t)}{t} 
\end{equation}
for $t<t_{\rm dec}$. The trivial single-velocity shell result that $R \propto t$ for $t<t_{\rm dec}$ is recovered in the limit $\alpha \to \infty$.

In contrast to the single-velocity shell scenario there is no well-defined ``collision'' time between the ejecta and jet forward shock. This is because fast-moving components of the ejecta will catch-up with the jet earlier than slower moving ejecta.
What therefore is the relevant timescale equivalent to $t_{\rm col}$?
This timescale is the one which demarcates the transition from forward-shock dynamics dominated by the jet versus one dominated by the ejecta. This occurs roughly once the blast-wave running into the external ISM becomes energetically dominated by the ejecta instead of the jet, i.e. once ejecta with initial velocity $\bar{v}_{\rm col}$, defined such that $E_{\rm ej}(\geq \bar{v}_{\rm col}) \equiv E_{\rm j}$, has caught up to the jet-forward shock. From equation~(\ref{eq:Appendix_Egeqv}) this yields
\begin{equation}
    \bar{v}_{\rm col} = v_{\rm ej} \left(\frac{E_{\rm j}}{E_{\rm ej}}\right)^{-1/\alpha} ,
\end{equation}
and making the simplifying assumption that the ejecta expands homologously within the jet-cavity (i.e. neglecting deceleration of ejecta components still contained within $R_{\rm j}$) --- the corresponding timescale is
\begin{equation}
    \bar{t}_{\rm col} = t_{\rm col}(\bar{v}_{\rm col})
    = t_{\rm col}(v_{\rm ej}) \left(\frac{E_{\rm j}}{E_{\rm ej}}\right)^{5/3\alpha} 
    =2.04 \left(\frac{E_{\rm j}}{E_{\rm ej}}\right)^{(\alpha+5)/3\alpha} t_{\rm dec}
    .
\end{equation}
In the above, $t_{\rm col}(v)$ is the standard single-velocity shell collision time (eq.~\ref{eq:tcol}) for the `bulk' of the ejecta travelling at $v=v_{\rm ej}$.

At times $t \ll \bar{t}_{\rm col}$ the forward shock with the external ISM follows the Sedov-Taylor solution with $E=E_{\rm j}$ (eqs.~\ref{eq:Rjvj}), while at times $t \gg \bar{t}_{\rm col}$ the ejecta energy sets the forward-shock dynamics that will follow equation~(\ref{eq:Appendix_Rv}) at $t<t_{\rm dec}$ and equation~(\ref{eq:Rjvj}) with $E=E_{\rm ej}$ at $t>t_{\rm dec}$.

Following \S\ref{sec:non-interacting}, the synchrotron luminosity of ejecta colliding into an unperturbed constant-density ISM evolve as (see also \citealt{Kathirgamaraju+16})
\begin{equation}
    L_{\nu,{\rm ej}}(t<t_{\rm dec}) \propto 
    \begin{cases}
    t^\frac{3(2\alpha-5p+7)}{2(\alpha+5)}; &{\rm else}
    \\
    t^\frac{3(2\alpha-p-1)}{2(\alpha+5)}; &{\rm DN ~ regime}
    \end{cases}
    .
\end{equation}{}
With these scalings, the time $\bar{t}_{\rm eq}$ at which the declining jet afterglow and rising ejecta radio-flare signals would equal one another (neglecting jet--ejecta interaction) is given by
\begin{equation}
    \bar{t}_{\rm eq} = \bar{t}_{\rm col}
    \begin{cases}
    \left[\frac{2(\alpha+5)}{5(\alpha+2)}\right]^\frac{5(\alpha+5)(5p-3)}{3\alpha(5p+3)}; &{\rm else}
    \\
    \left[\frac{2(\alpha+5)}{5(\alpha+2)}\right]^\frac{5(\alpha+5)(p+5)}{3\alpha(p+11)}; &{\rm DN ~ regime}
    \end{cases}
    .
\end{equation}
The term in brackets represents the ratio of jet and ejecta forward-shock velocities at time $\bar{t}_{\rm col}$ as determined by equations~(\ref{eq:Rjvj},\ref{eq:Appendix_Rv}).
Once again, it can be verified that this result reduces to the single-velocity shell case (eq.~\ref{eq:teq}) for $\alpha \to \infty$, as expected.

\section{Wind Ambient-Density}
\label{sec:Appendix_winddensity}

In the main text we discussed results for the case where the ambient medium surrounding the progenitor is a constant density ISM. Here we extend these results to the case of a density profile that decreases as $r^{-2}$, relevant to systems where the progenitor may have launched strong stellar winds.
The ambient density in this scenario can be written as
\begin{equation}
    \rho = A r^{-2} = 5 \times 10^{11} \, {\rm g \, cm}^{-1} \, A_\star r^{-2},
\end{equation}
where 
$A = \dot{M}_{\rm w}/4\pi v_{\rm w}$
is normalized to $A = 5 \times 10^{11} A_\star \, {\rm g \, cm}^{-1}$ appropriate for a characteristic mass-loss rate $\dot{M}_{\rm w} = 10^{-5} M_\odot \, {\rm yr}^{-1}$ and wind velocity $v_{\rm w} = 1,000 \, {\rm km \, s}^{-1}$.

The Sedov-Taylor solution for the jet forward-shock in a wind ambient density dictates
\begin{equation}
    R_{\rm j} = \left(2.1\frac{E_{\rm j}}{A}\right)^{1/3} t^{2/3}; ~~~~~~~~~~
    v_{\rm j} = \frac{2}{3} \frac{R_{\rm j}(t)}{t} .
\end{equation}
Assuming the ejecta expands within the jet-shocked ISM uninhibited (coasting at constant velocity) then the above implies that the collision time at which the ejecta catches-up with the jet is
\begin{equation}
    t_{\rm col} = 2.1 \left(\frac{E}{A}\right) v_{\rm ej}^{-3}
    \approx 49 \, {\rm yr} \, A_\star^{-1} E_{{\rm j},49} \beta_{{\rm ej},-1}^{-3} .
\end{equation}
Furthermore, the Sedov-Taylor solution interior to the the jet forward-shock ($r<R_{\rm j}$) in the wind density profile is simply $\rho \propto (r/R_{\rm j}) \rho(R_{\rm j})$ and $v \propto r/t$. The density at the shock front is $\rho(R_{\rm j}) \propto R_{\rm j}^{-2} \propto t^{-4/3}$. Using these results along with eq.~(\ref{eq:Lnu_ej_tscaling}), we find that the optically-thin synchrotron light-curve produced by the ejecta expanding within the jet-shocked medium scales as
\begin{equation}
    L_{\nu,{\rm ej}}(t<t_{\rm col}) \propto t^\frac{7-p}{4}
\end{equation}
prior to $t_{\rm col}$.
At later times the light-curve converges to the $L_{\nu,{\rm ej}} \propto t^{-(p-1)/2}$ behavior for a constant velocity ejecta in an $r^{-2}$ density profile. Note that contrary to the constant-density ambient medium case, the light-curve is declining as a function of time, even prior to $t=t_{\rm dec}$ (although synchrotron self-absorption would cause the light-curve to rise at very early times).

\begin{figure}
    \centering
    \includegraphics[width=0.45\textwidth]{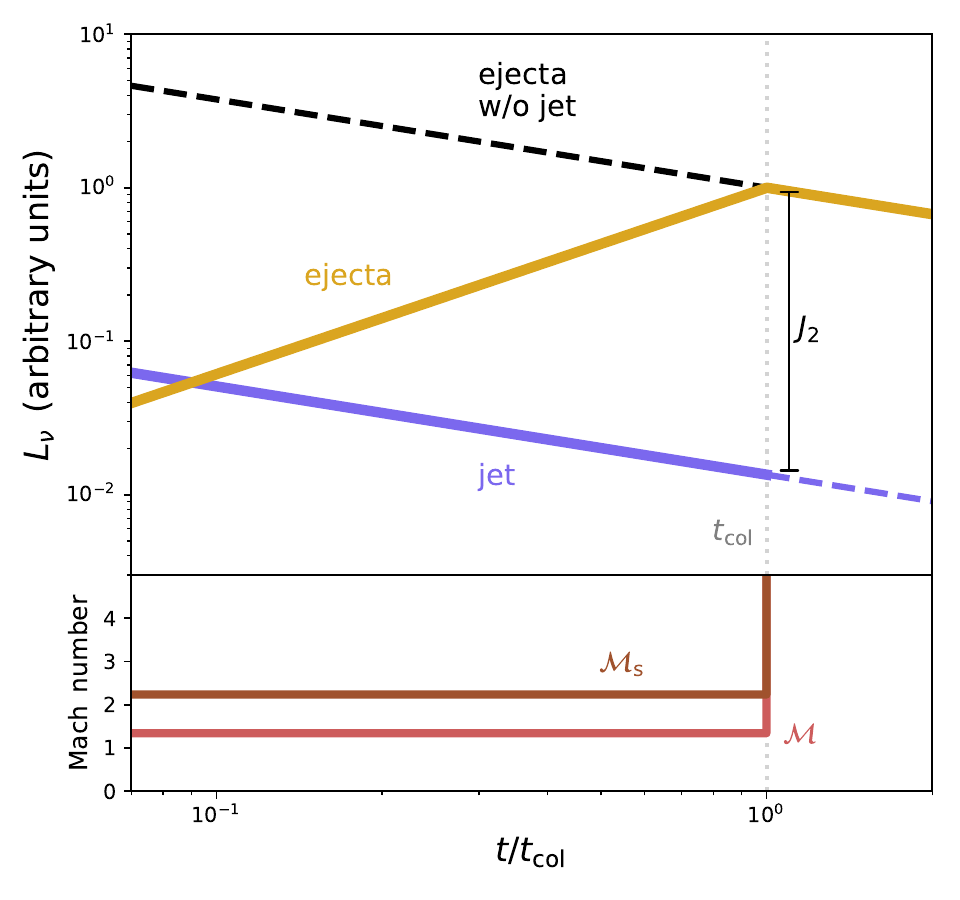}
    \caption{
    Same as Fig.~\ref{fig:numerical_results} but for a wind external density profile ($\rho \propto r^{-2}$). This is calculated assuming $p=2.15$ and that both the jet and ejecta are within the deep-Newtonian regime. The light-curve temporal evolution can be fully-solved analytically in this scenario, see text for further details.
    }
    \label{fig:wind}
\end{figure}

We can also calculate the jump in the light-curve immediately before and after $t=t_{\rm col}$ (the equivalent of eq.~\ref{eq:J1}). For a wind density profile and adiabatic index $g=5/3$ the shock compression ratio (ratio of upstream density ahead of the ejecta immediately before $t_{\rm col}$ and immediately afterwards) is $4$, and the gas velocity at the jet-wind forward shock is $v(R_{\rm j}) = r / 2t$. Thus, the jump factor can be calculated from eq.~(\ref{eq:Lnu_ej_tscaling}) to be
\begin{equation}
    J_1 = 
    \begin{cases}
    2^{-2(p-2)} ; &t_{\rm col}<t_{\rm DN}
    \\
    1 ; &t_{\rm col}>t_{\rm DN}
    \end{cases}
    .
\end{equation}
Interestingly, for the fiducial scenario where $t_{\rm col} > t_{\rm DN}$ (and also for $p=2$ in the $t>t_{\rm DN}$ regime) we find that $J_1=1$ and there is no jump in the luminosity for a wind ambient medium.
Similarly, the ratio of ejecta radio flare and jet afterglow luminosities at time $t=t_{\rm col}$ is given as
\begin{equation}
    J_2 =
    \begin{cases}{}
    \left(\frac{3}{10}\right)^{-\frac{5p-3}{2}}; &t_{\rm col}<t_{\rm DN}
    \\
    \left(\frac{3}{10}\right)^{-\frac{p+5}{2}}; &t_{\rm col}>t_{\rm DN}
    \end{cases}
    ,
\end{equation}
analogous to eq.~(\ref{eq:J2}) for the constant density ISM case.

A major caveat to the above estimates is the extremely low Mach number in the wind scenario. Similar to eq.~(\ref{eq:Mach}) we find that for a wind ambient density the Mach number of the ejecta with respect to the jet-shocked wind (and an adiabatic index $g=5/3$) is 
$\mathcal{M} = 3/\sqrt{5} \approx 1.34$. Furthermore, the Sedov-Taylor interior solution in this regime implies $c_{\rm s} \propto r/t$ and therefore $\mathcal{M} = const.$ as a function of time.
The corresponding sonic Mach number $\mathcal{M}_{\rm s} \approx 2.24$ is thus below the critical value identified by \cite{Kang+19} where particle acceleration is ineffective ($\mathcal{M}_{\rm s} \lesssim 2.3$). This may imply that radio emission by the ejecta is entirely quenched at $t<t_{\rm col}$, but again we caution that further work is needed to clearly address this issue. In particular, and as discussed in \S\ref{sec:interaction}, the presence of relic high-energy electrons accelerated at the jet-wind forward shock may contribute to alleviating particle-acceleration inefficiencies.

The sphericization timescale for the jet can be similarly generalized from a constant-density to a wind ambient medium. The analog of eq.~(\ref{eq:tsph}) is
\begin{equation}
    t_{\rm sph} \gtrsim t_{\rm NR} \sim \left(\frac{E_{\rm j,iso}}{4\pi A c^3} \right)
    \approx 0.37 \, {\rm yr} \, A_\star^{-1} E_{{\rm j},49} \theta_{{\rm j},-1}^{-2} .
\end{equation}
From the above equations we find that the assumption of spherical symmetry is valid at $t=t_{\rm col}$ (analogous to eq.~\ref{eq:sph_assumption})
if
\begin{equation}
    \theta_{\rm j} \gtrsim 0.015 \, {\rm rad} \, \left(\frac{f_{\rm sph}}{3}\right)^{1/2} \beta_{{\rm ej},-1}^{3/2} .
\end{equation}
We normalized $f_{\rm sph}$ in the above to a slightly lower value than in eq.~(\ref{eq:sph_assumption}), roughly consistent with numerical simulations, but even for somewhat higher values $f_{\rm sph} \sim 10$ the assumption of spherical symmetry is more easily satisfied in a wind environment (see Fig.~\ref{fig:spherical_assumption}).

The decceleration timescale of the ejecta is
\begin{equation}
    t_{\rm dec} = \left(\frac{E_{\rm ej}}{4\pi A v_{\rm ej}^3} \right)
    \approx 190 \, {\rm yr} \, A_\star^{-1} E_{{\rm ej},51} \beta_{{\rm ej},-1}^{-3} .
\end{equation}
At later times the synchrotron light-curve evolves temporally as
\begin{equation}
    L_{\nu,{\rm ej}}(t>t_{\rm dec}) \propto
    \begin{cases}
    t^{-\frac{7p-5}{6}}; &t_{\rm col}<t_{\rm DN}
    \\
    t^{-\frac{p-1}{2}}; &t_{\rm col}>t_{\rm DN}
    \end{cases}
    .
\end{equation}
Therefore, in the deep-Newtonian regime there is no break in the light-curve at $t_{\rm dec}$.

The estimates above apply as long as the jet and ejecta propagate within the wind-zone, so that $\rho \propto r^{-2}$. In a realistic setting, this profile is expected to change at distances larger than the wind termination shock,
$R_{\rm w} \approx 9 \, {\rm pc} \, A_\star^{3/10} n_0^{-3/10} v_{{\rm w},8}^{2/5} (t / {\rm Myr})^{2/5}$ \citep{Weaver+77}.
This should be compared to the radii of relevance in our current work $\sim R_{\rm col} = v_{\rm ej} t_{\rm col} \approx 1.5 \, {\rm pc} \, A_\star^{-1} E_{{\rm j},49} \beta_{{\rm ej},-1}^{-2}$. As long as $R_{\rm col} < R_{\rm w}$ the assumption of an $\rho \propto r^{-2}$ wind environment is self-consistent. Otherwise, the dynamics and resulting light-curve will be altered \citep[e.g.][]{vanMarle+06}.


\bsp	
\label{lastpage}
\end{document}